\documentclass[3p,onecolumn,11pt,sort&compress]{elsarticle}

\usepackage{graphicx}
\usepackage[american]{babel}
\usepackage[utf8]{inputenc}
\usepackage[T1]{fontenc}
\usepackage{latexsym,amsmath,amssymb}
\usepackage{booktabs}

\usepackage{float}
\usepackage{placeins}
\usepackage{url}

\allowdisplaybreaks
\usepackage[sc]{mathpazo}
\usepackage{caption}
\usepackage[ruled,vlined,noend]{algorithm2e}

\usepackage{booktabs}
\usepackage{tabularx}
\usepackage{threeparttable}
\usepackage{siunitx}
\usepackage[gen]{eurosym}

\usepackage{color}

\usepackage{cleveref}
\usepackage[hidelinks]{hyperref}

\usepackage{tikz}
\usepackage[europeanresistors,americaninductors]{circuitikz}
\usepackage{adjustbox}
\usepackage{booktabs}
\usepackage{lipsum}

\newcommand{\fref}[1]{Figure~\ref{#1}}

\renewcommand{\eqref}[1]{Equation~(\ref{#1})}

\usepackage{nomencl}
\usepackage{framed}
\makenomenclature

\begin{document}

\begin{frontmatter}

\title{A validated high-resolution hydro power time-series model for energy systems analysis}

\author[label1]{Hailiang Liu}
\ead{HLL@eng.au.dk}
\author[label1]{Gorm Bruun Andresen}
\author[label2]{Tom Brown}
\author[label1]{Martin Greiner}

\address[label1]{Department of Engineering, Aarhus University, Inge Lehmanns Gade 10, 8000 Aarhus C,  Denmark}
\address[label2]{Institute for Automation and Applied Informatics, Karlsruhe Institute of Technology, 76344 Eggenstein-Leopoldshafen, Germany}

\journal{MethodsX}

\begin{abstract}

We expand the renewable technology model palette and present a validated high resolution hydro power time series model for energy systems analysis. Among the popular renewables, hydroelectricity shows unique storage-like flexibility, which is particularly important given the high variability of wind and solar power. Often limited by data availability or computational performance, a high resolution, globally applicable and validated hydro power time series model has not been available. For a demonstration, we focus on 41 Chinese reservoir-based hydro stations as a demo, determine their upstream basin areas, estimate their inflow based on gridded surface runoff data and validate their daily inflow time series in terms of both flow volume and potential power generation. Furthermore, we showcase an application of these time series with hydro cascades in energy system long term investment planning. Our method's novelty lies in:
\begin{itemize}
	
\item it is based on highly resolved spatial-temporal datasets;

\item both data and algorithms used here are globally applicable; 

\item it includes a hydro cascade model that can be integrated into energy system simulations.
	
\end{itemize}

\end{abstract}

\begin{keyword}

hydro power \sep
validation \sep
reanalysis \sep
energy systems \sep
China

\end{keyword}

\end{frontmatter}

\clearpage

\section{Introduction}

Research on the decarbonizing transformation of  energy systems calls for robust high-resolution renewable energy generation data and models. Among the most popular renewables, there have been state-of-the-art wind and solar power time series models, validated and applied for various countries \cite{pfenninger2016long,staffell2016using,andresen2015validation,LIU2018534}. However, hydroelectricity, as another important sustainable power source, has not been properly represented in energy systems modeling. In countries such as Norway, Iceland, Switzerland, Brazil, Canada and China, hydro power supplies the largest share among renewables or even all energy sources \cite{Bilgili2015Sep}. Furthermore, in the decarbonization of the energy sectors, a high penetration of variable renewable power is expected \cite{kozarcanin2018climate,zhu2018impact,Liu2018Oct}. In these systems, the storage-like flexibility of hydro power can be pivotal. 

Hydro power integrating with other renewables has been considered for off-grid remote sites, regional or continental long term investment planning. Limited by either data availability \cite{Thangavelu2015Sep} or computational power \cite{Li2016Apr}, their representation of hydroelectricity is restricted to a handful of reservoir stations with low temporal resolution \cite{Khan2018Jan}. These time series from the TSOs \cite{Schmidt2016Jan} usually are in several discontinuous periods, and cascade coordinations are masked.

We present a new validated high resolution hydro power time series model, designed for energy systems analysis. In a nut shell, we focus on 41 large-scale reservoir-based hydro stations in China, determine their corresponding upstream basin areas, estimate their inflow based on gridded surface runoff data from CFSR \cite{cisl_rda_ds094.1} and calculate their daily inflow time series in terms of both flow volume and potential power generation. To our knowledge, no high resolution hydroelectricity generation time series have been modeled or validated before.

\section{Upstream basins determination}

Wind turbines and solar PVs' power generation depends on instantaneous local wind speed and solar radiation. Their power output does not depend on the weather conditions far away. For hydroelectricity, it is a different story. In fact, the vast majority of the reservoir inflow, whose kinetic energy is converted to electricity at the hydro dam, is not from the precipitation or snow-melt nearby. 

\begin{figure*}[h!]
	\centering
	\includegraphics[width=\linewidth]{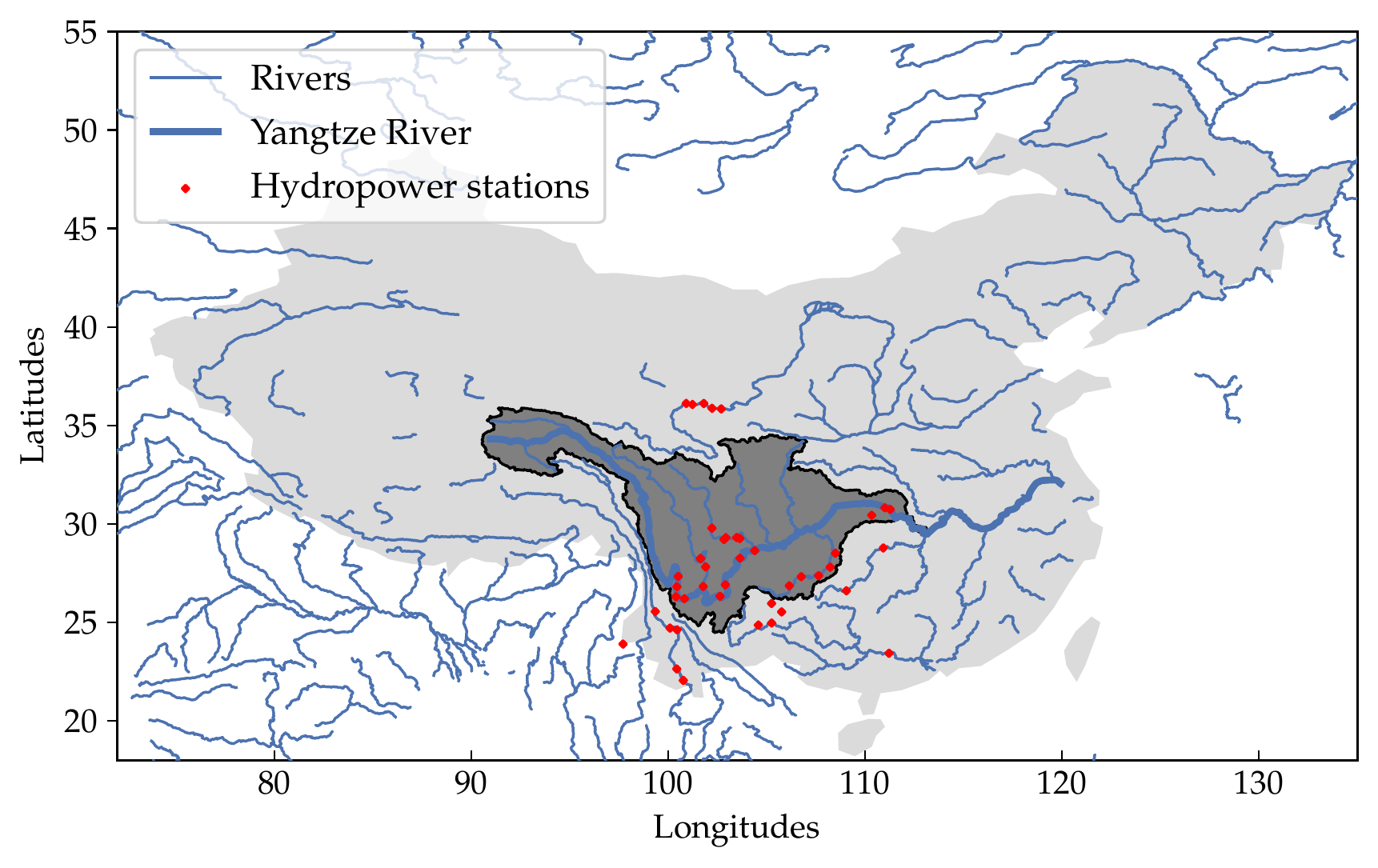}
	\caption{The 41 largest hydro stations spanning major rivers in the southwest. Most dams are part of a cascade on the same river. The dark grey area represents the largest basin for Yangtze River.}
	\label{dams}
\end{figure*}

We only consider the 41 largest reservoir hydro stations in China (\fref{dams}). Run-of-river, whose generation varies upon instantaneous inflow is not included here. Spanning over major rivers, hydro reservoirs' inflow is highly seasonal, and they depend on the precipitation in the upstream areas. The inflows are mainly surface runoffs routed into their upstream river networks, which eventually accumulate in the reservoirs. Usually, river basins are well-defined and documented from source to mouth. However, only basin areas which lie upstream of the hydro stations affect the reservoir inflows, and often the stations do not lie at basin borders.

\begin{figure*}[h!]
	\centering
	\includegraphics[width=0.495\linewidth]{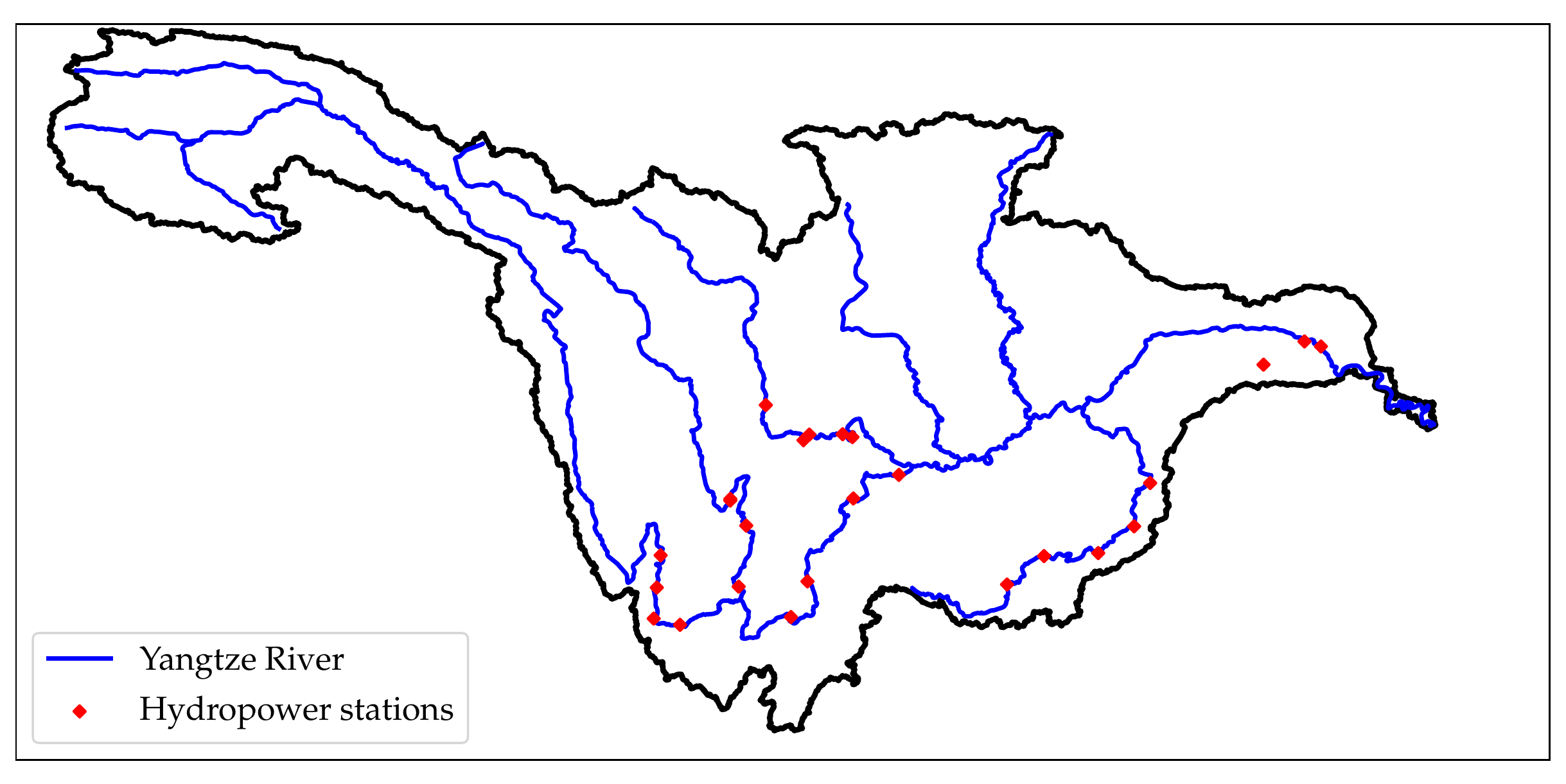}
	\includegraphics[width=0.495\linewidth]{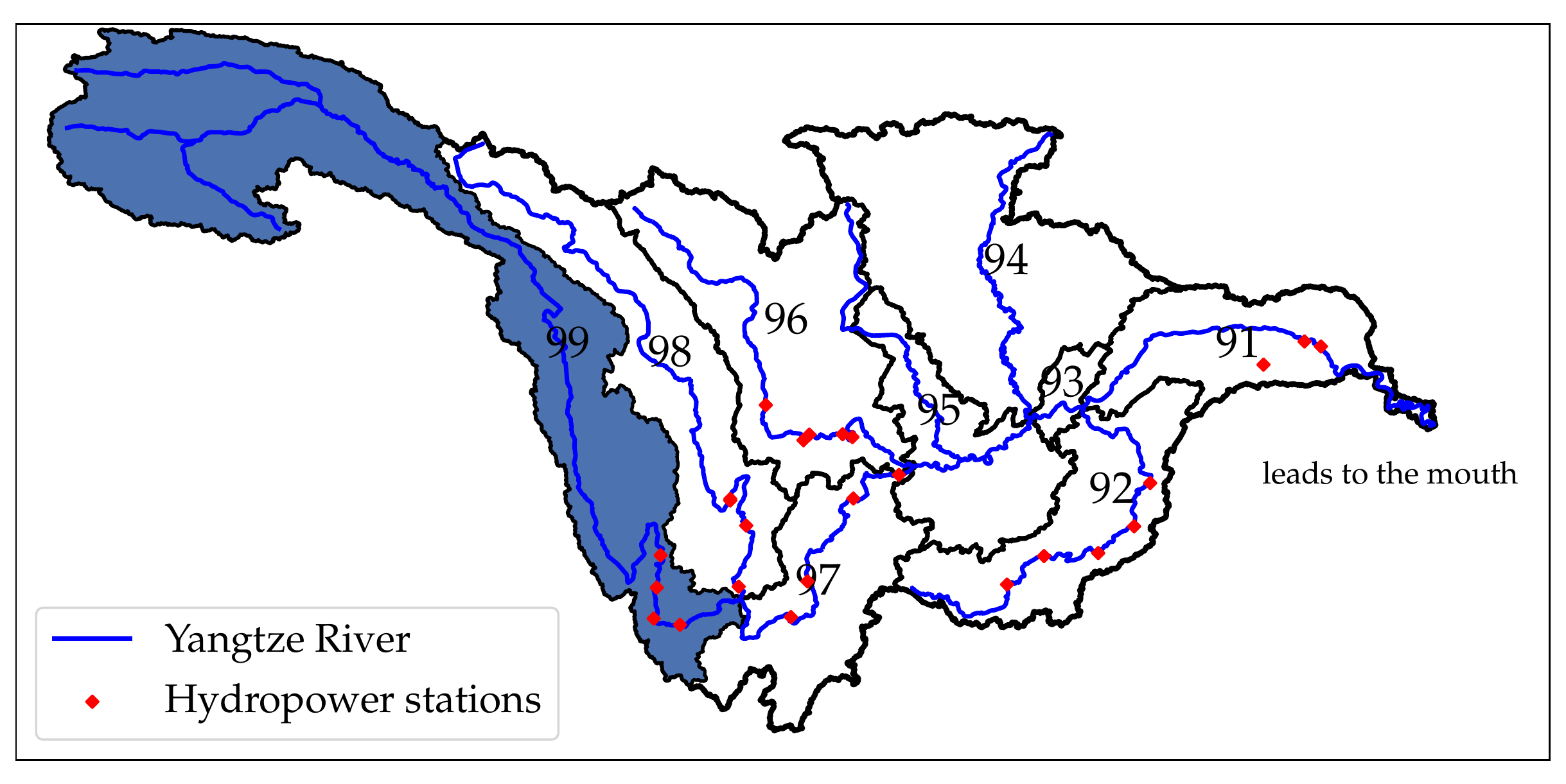}
	\includegraphics[width=0.495\linewidth]{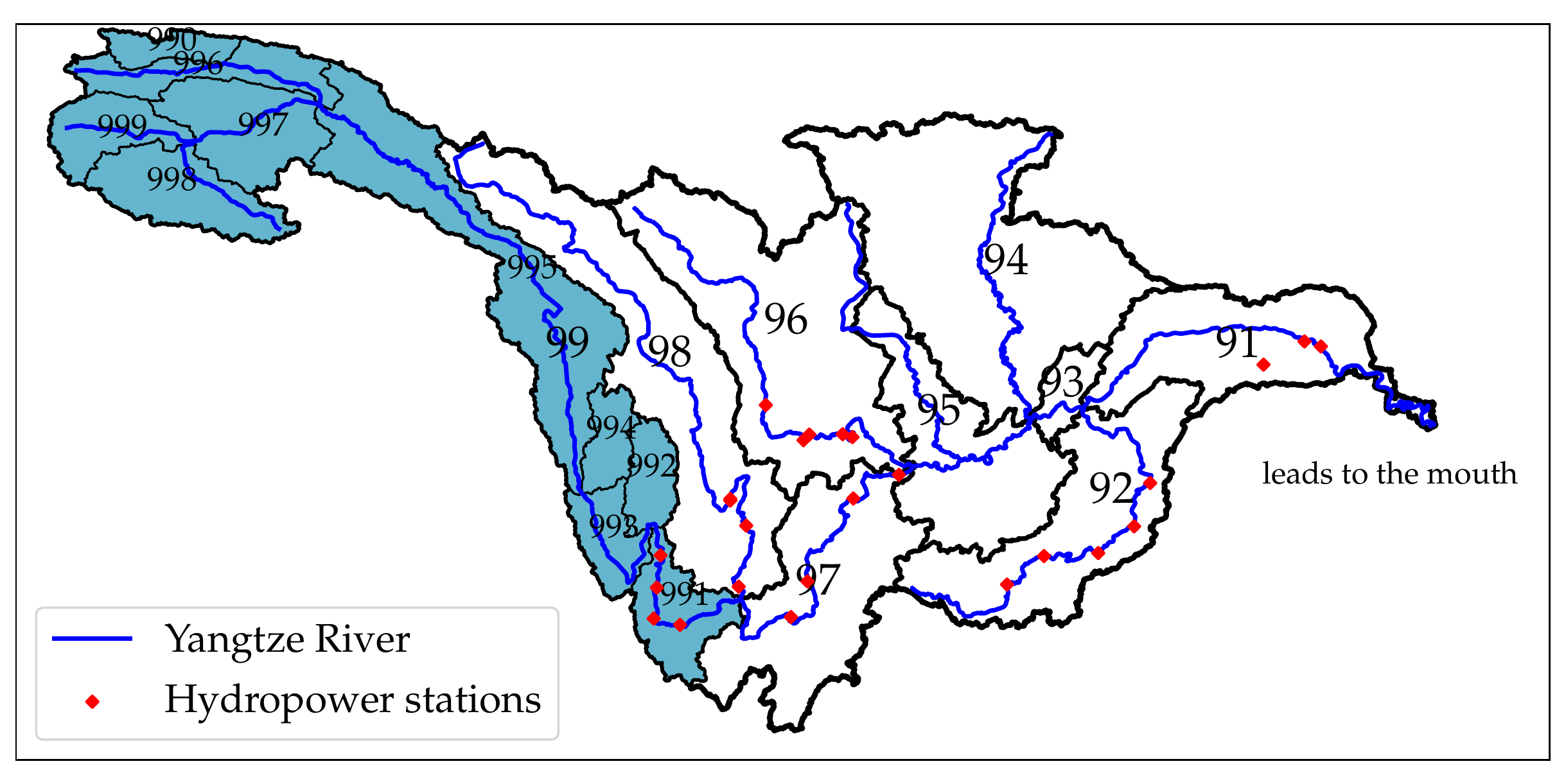}
	\includegraphics[width=0.495\linewidth]{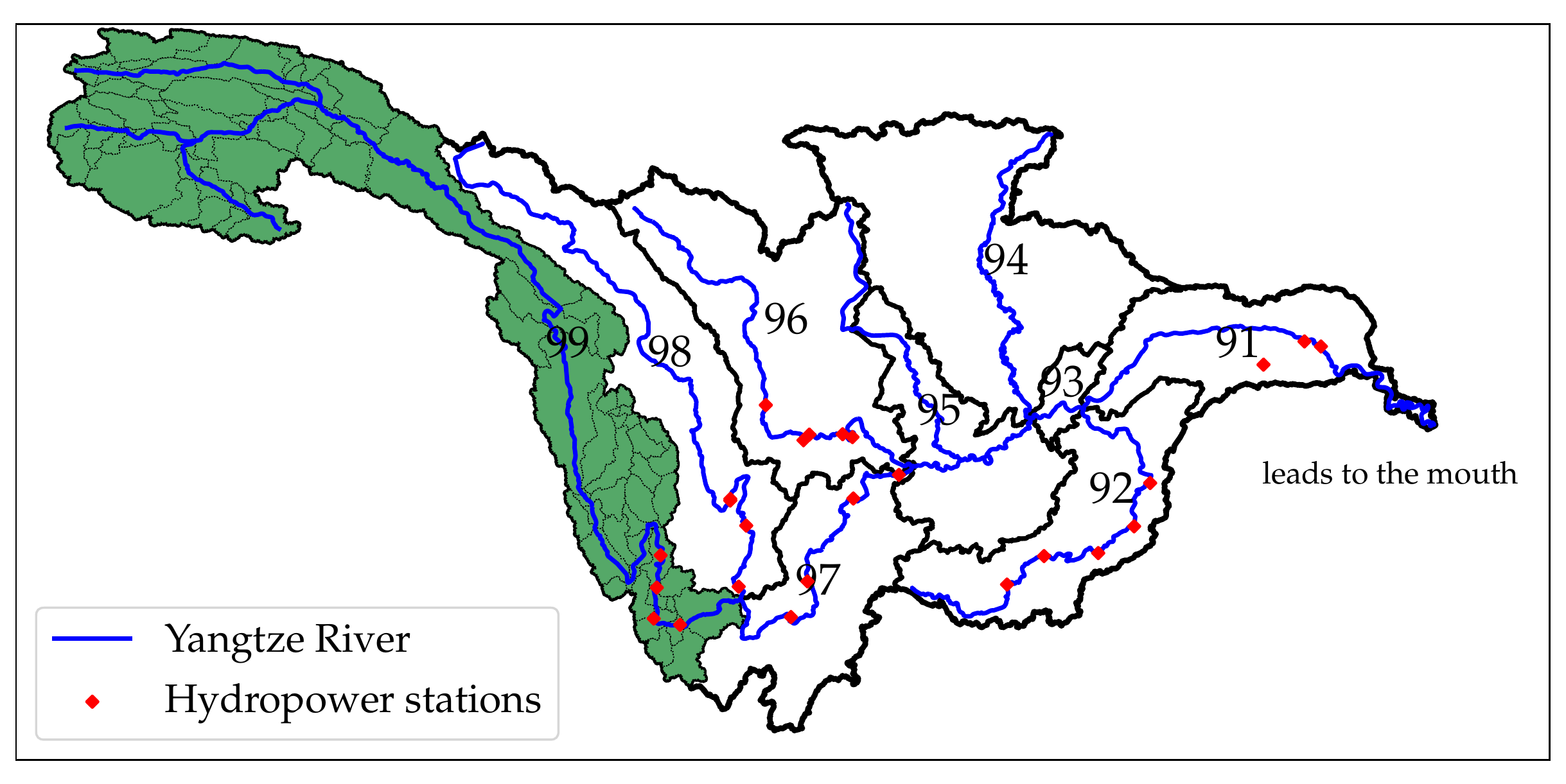}
	\caption{Considering the fine spatial resolution in our model, we use basin levels 4, 5, 6 and 7 of the HydroBASINS dataset. The larger the number, the finer the resolution is. There are 22 hydro stations on the main stem or tributaries of Yangtze River. The river network is only drawn to verify the basin delineations. On level 4 (a), the basin's Pfafstetter code is 4349. Its odd last digit means that the basin lies on the main stream of the river. It is divided into 9 higher level basins, coded as 43491, 43492 and so on. In b, we only show the last two digits for better visualization. The most upstream basin is indexed 43499 and colored blue. It is further divided into 9 higher level basins shown in c, numbered 434991 through 434999, and in the figure they are labeled 991 through 999. Four hydro stations fall in the basin 434991, and this means their reservoir inflow comes from the runoff in basins 434991 through 434999. To raise the accuracy higher, level 7 basins in d show distinct delineations among the four station close to each other. }
	\label{levels_of_basins}
\end{figure*}

\begin{figure*}[h!]
	\centering
	\includegraphics[width=\linewidth]{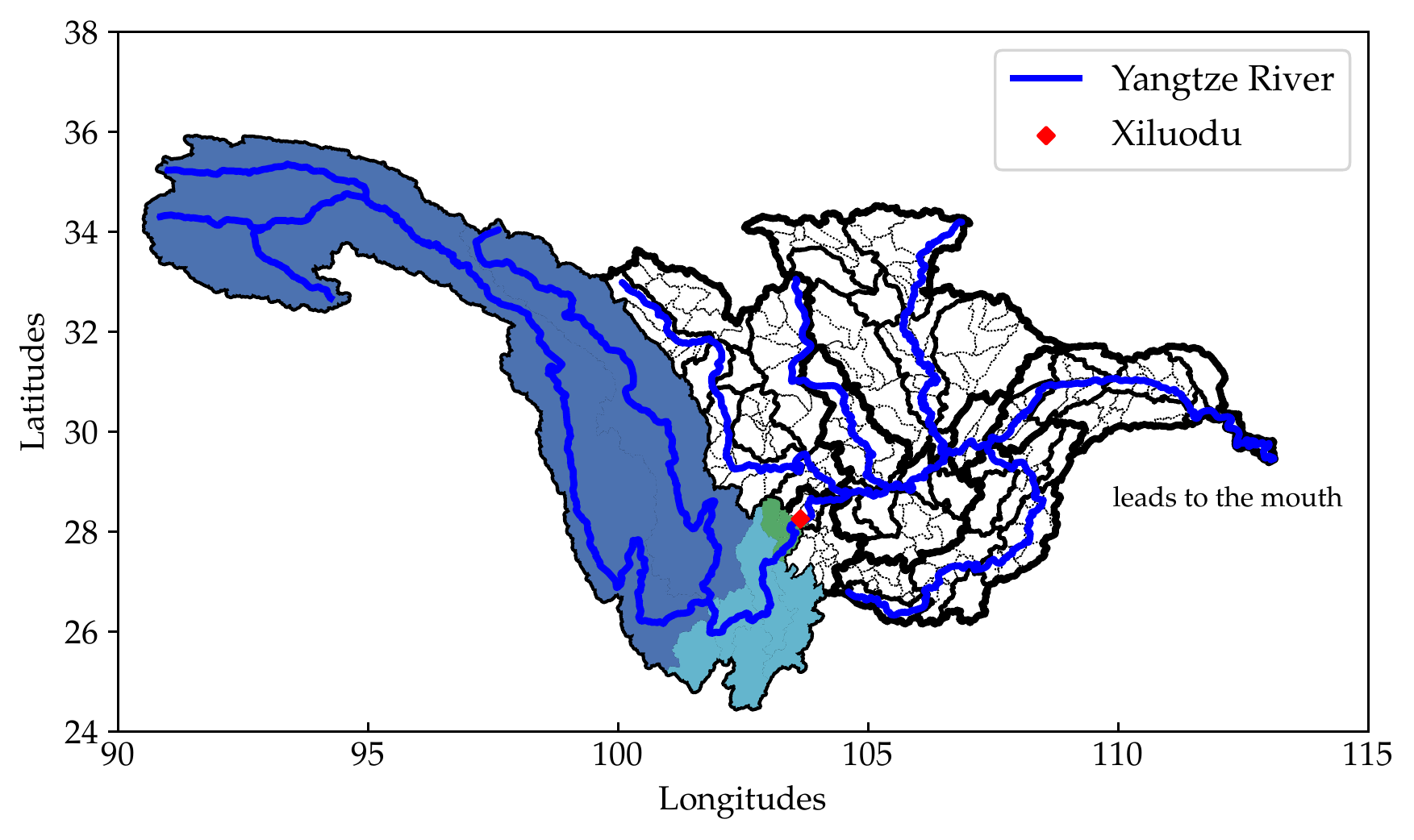}
	\caption{Upstream basins of hydro reservoirs are determined using the HydroBASINS dataset \cite{lehner2013global} and the Pfafstetter Coding System \cite{verdin1999topological}, the procedure of which is described in detail in Algorithm \ref{code_basin_determination}. Thick solid lines represent higher level basin boundaries, and thin dashed lines enclose lower level basins. Hydro station \textit{Xiluodu} collects surface runoff from the colored areas, and different colors denote various basin levels.}
	\label{Xiluodu}
\end{figure*}

The HydroBASINS dataset \cite{lehner2013global}, is a set of polygon layers that show basin boundaries and sub-basin delineations at a global scale. It provides a seamless coverage of consistently sized and hierarchically nested sub-basins at different scales (from tens to millions of square kilometers). It is indexed using the Pfafstetter coding scheme \cite{verdin1999topological} that allows for analysis such as up- and downstream basin connectivity. We used the three most important features of the Pfafstetter scheme to determine the hydro stations upstream basins: Odd digits denote basin segments on the main stem, even digits denote tributaries of the main stem. At each level, higher digits denote upper stream segments. A basin's Pfafstetter code with lower resolution is exactly the same as its finer scale basin taking out the last digit \cite{verdin1999topological}. This is further illustrated in \fref{levels_of_basins}.

To determine a dam's upstream basins, explained in Algorithm \ref{code_basin_determination}, we use a nested loop to search through the HydroBASINS polygon series. First, the geographical coordinates of the dam gives the level 7 basin it lies in. If its Pfafstetter code $ PFcode $ is an even number, meaning it is a tributary, then it is this dam's only upstream basin. Otherwise, it is a main stem, and there are more basins upstream. They can be found by returning all level 7 basins with the same Pfafstetter code but the last digit. Then, we pop the last digit of $ PFcode $ and if the new $ PFcode $ is an even number then we can finish. Otherwise, we search for all level 6 basins with the same Pfafstetter code but the last digit. Finally, the same procedure goes for level 5 basins. An example is shown in \fref{Xiluodu}. Note that the rivers are only drawn to verify the basin delineations and they are not used in the upstream determination.

\begin{algorithm*}[t!]
	\SetAlgoLined
	\DontPrintSemicolon
	\KwData{Three datasets consisting three levels of basins $ basins7, basins6, basins5 $}
	\KwResult{List of $ upstream\_basins $ that lie upstream of the hydro dam $ Dam $}
	$ coordinates \gets $ coordinates of $ Dam $\;
	\For{$ Basin7 \in basins7 $}{
		\If{$ coordinates $ is in $ Basin7 $}{
			$ PFcode  \gets pfcode(Basin7) $\;
		}	
	}
	append $PFcode$ to $upstream\_basins $\;
	\eIf{$ PFcode  $ is an even number, meaning it's a tributary}
	{ 
		Finish\;} (it's a main stem)
	{\For{$ Basin7 \in basins7 $}{
			$ p \gets pfcode(Basin7) $ \;
			{\If{$p > PFcode  $ and they are the same but last digit}
				{append $p$ to $upstream\_basins$\;}}	
		}
		$ PFcode \gets $ all but last digit of $ PFcode $\;
		\eIf{$PFcode$ is an even number, meaning it's a tributary}{Finish\;} (it's a main stem)
		{
			\For{$ Basin6 \in basins6 $}
			{$ p \gets pfcode(Basin6) $\;
				\If{$ p > PFcode  $ and they are the same but last digit}
				{append $p$ to $upstream\_basins$\;}
			}
			$ PFcode \gets $ all but last digit of $ PFcode $\;
			\eIf{$PFcode$ is an even number, meaning it's a tributary}{Finish\;} (it's a main stem)
			{
				\For{$ Basin5 \in basins5 $}
				{$ p \gets pfcode(Basin5) $\;
					\If{$ p > PFcode  $ and they are the same but last digit}
					{ append $p$ to $upstream\_basins$ \;}
				}
			}
		}
		
	}
	\caption{Determination of a dam's upstream basins}
	\label{code_basin_determination}
\end{algorithm*}

\section{Reservoir inflow calibration}

Reservoirs' inflows are highly correlated to the surface runoff over its upstream basins. Runoff data is obtained from a global 38-year-long high-resolution weather dataset called CFSR (Climate Forecast System Reanalysis) from NCEP (American National Centers for Environmental Prediction) \cite{cisl_rda_ds094.1}. It is a reanalysis dataset, which means that it assimilates weather observations into a numerical weather prediction model. The weather model is a global, state-of-the-art, coupled atmospheric and oceanic model. Surface runoff is structured as a globally 0.3x0.3 degree gridded, hourly time series in the unit of $ kg/m^2 $.

\begin{figure*}[h!]
	\centering
	\includegraphics[width=\linewidth]{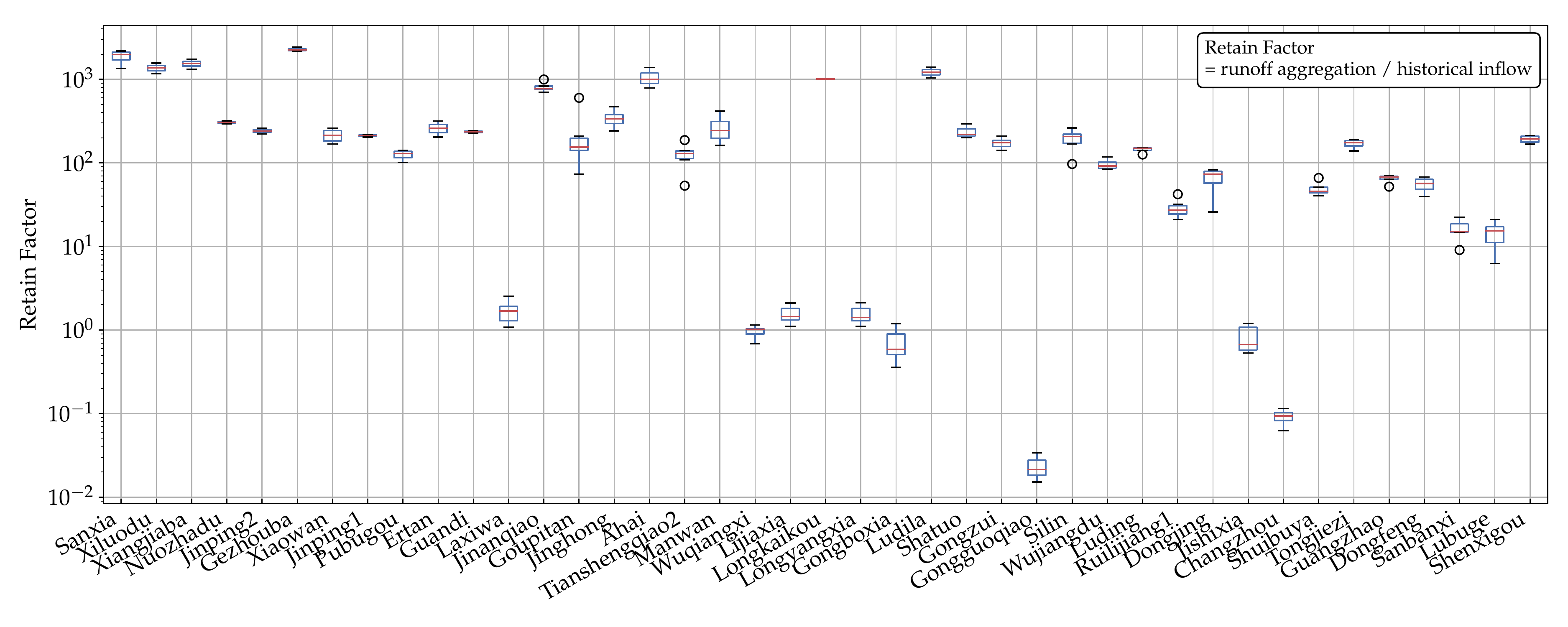}
	\caption{Retain factors for the 41 reservoirs. Several small reservoirs show values smaller than one due to their small upstream areas.}
	\label{retainfactor}
\end{figure*}

Surface runoff in the upstream basins are aggregated and calibrated against historical yearly reservoir inflow measurements \cite{Almanac} from 2009 to 2015, to account for evaporation, transpiration, irrigation, groundwater infiltration or runoff movement. Their ratio formulates an empirical parameter \textit{retain factor} for each reservoir, shown in \fref{retainfactor}. The results show that retain factors vary among reservoirs, and more importantly, for most this parameter shows minimal abnormality in the 7 years time. Therefore, we assume that the retain factors remain the same for years 1979 through 2016 in the inflow calculations.

This time series is also made to account for the delay of runoff from upstream locations to the reservoirs, with an assumed flow speed of 1 $m/s$ \cite{yamazaki2009deriving}. Specifically, a reservoir's total upstream basins are approximated as a square of the same area, the length of whose diagonal is assumed to be the typical travel distance for the runoff. The delays turn out to be ranging from 1 day to 2 weeks.

\section{Hydro power validation}

Hydro stations' power production per unit water depends on their head heights. It is very important to get as much head as possible, as more head means more power and therefore higher return on investment. The head, usually expressed in meters, is the elevation difference between the head (reservoir) water level and the tailwater (downstream) level. Here, the heads are approximated by dividing yearly water inflows by power generation in the same year, and averaged over 2009-2015  \cite{Almanac}. 

\begin{figure*}[h!]
	\centering
	\includegraphics[width=\linewidth]{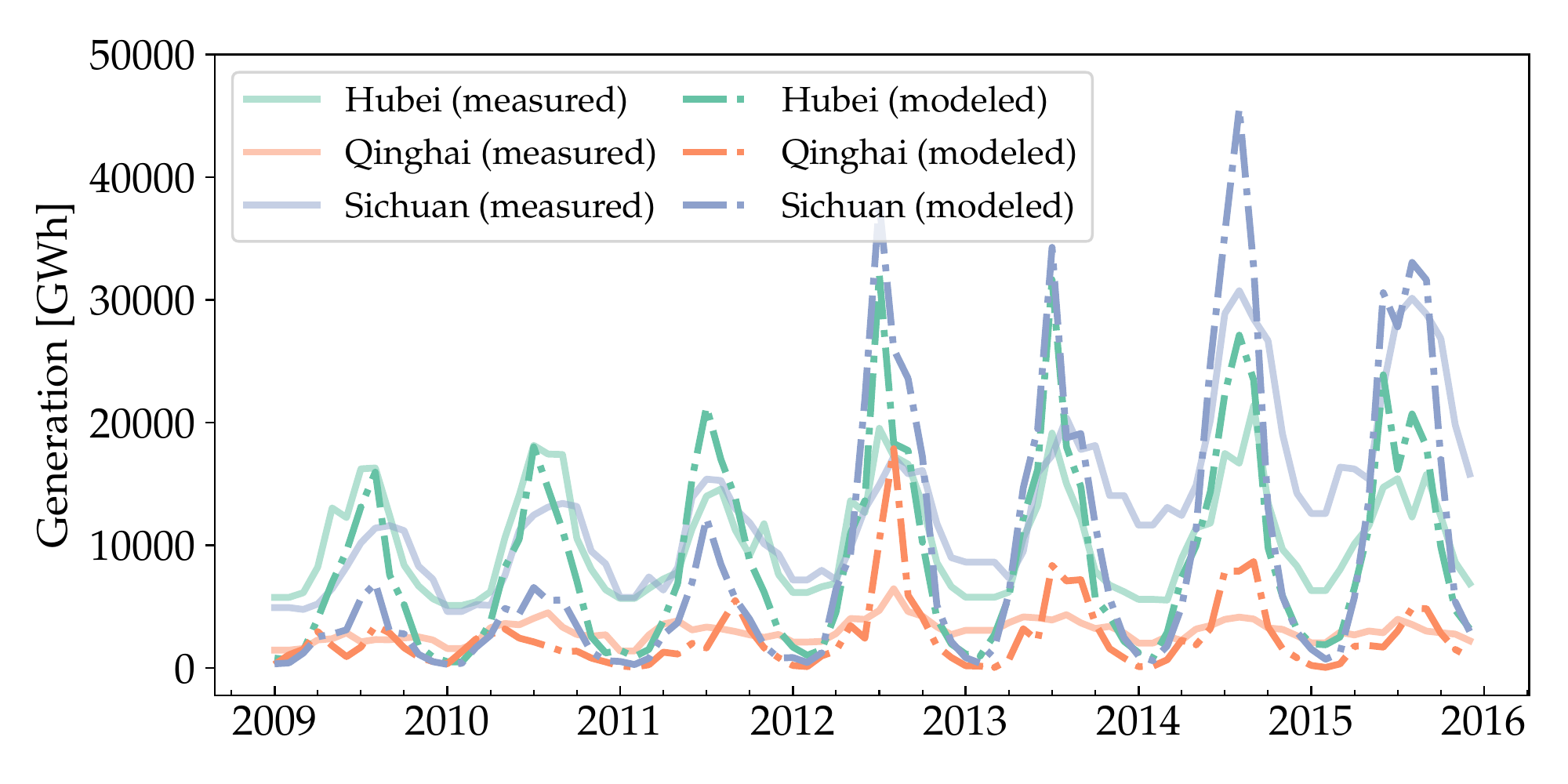}
	\includegraphics[width=\linewidth]{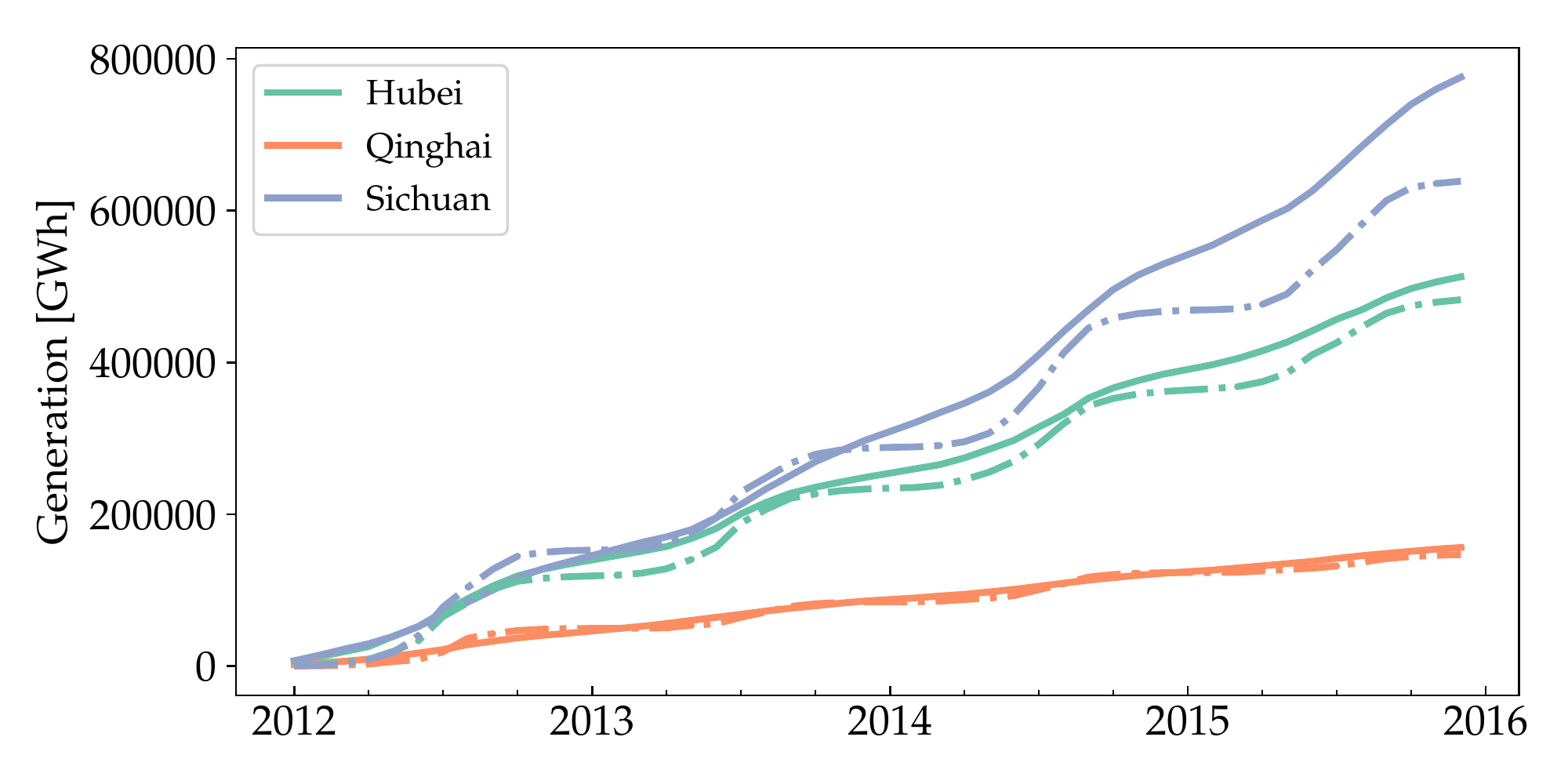}
	\caption{Sequential (top) and cumulative (bottom) measured and modeled hydro power generation for the provinces Hubei, Qinghai and Sichuan.}
	\label{validation}
\end{figure*}

Due to data availability, we are able to compare monthly hydroelectricity production on provincial level. \fref{validation} shows the measured and modeled power generation in the provinces Hubei, Qinghai and Sichuan. The stations in Hubei and Qinghai were in place before 2009 and are all included in the 41 stations under study here. In Sichuan, however, several stations like Xiluodu, Xiangjiaba and Jinping2 were only put in operation from 2012, which explains the increased power generation after that year. It is evident that, the model resembles historical long term power production patterns, while it can not capture the station operations such as saving water for low inflow seasons. This can be solved in energy systems models, since the station controls can be optimized together with other generators and storage units. We provide an example in Section \ref{cascades}.

\section{Daily inflow time series}

\begin{figure*}[h!]
	\centering
	\includegraphics[width=\linewidth]{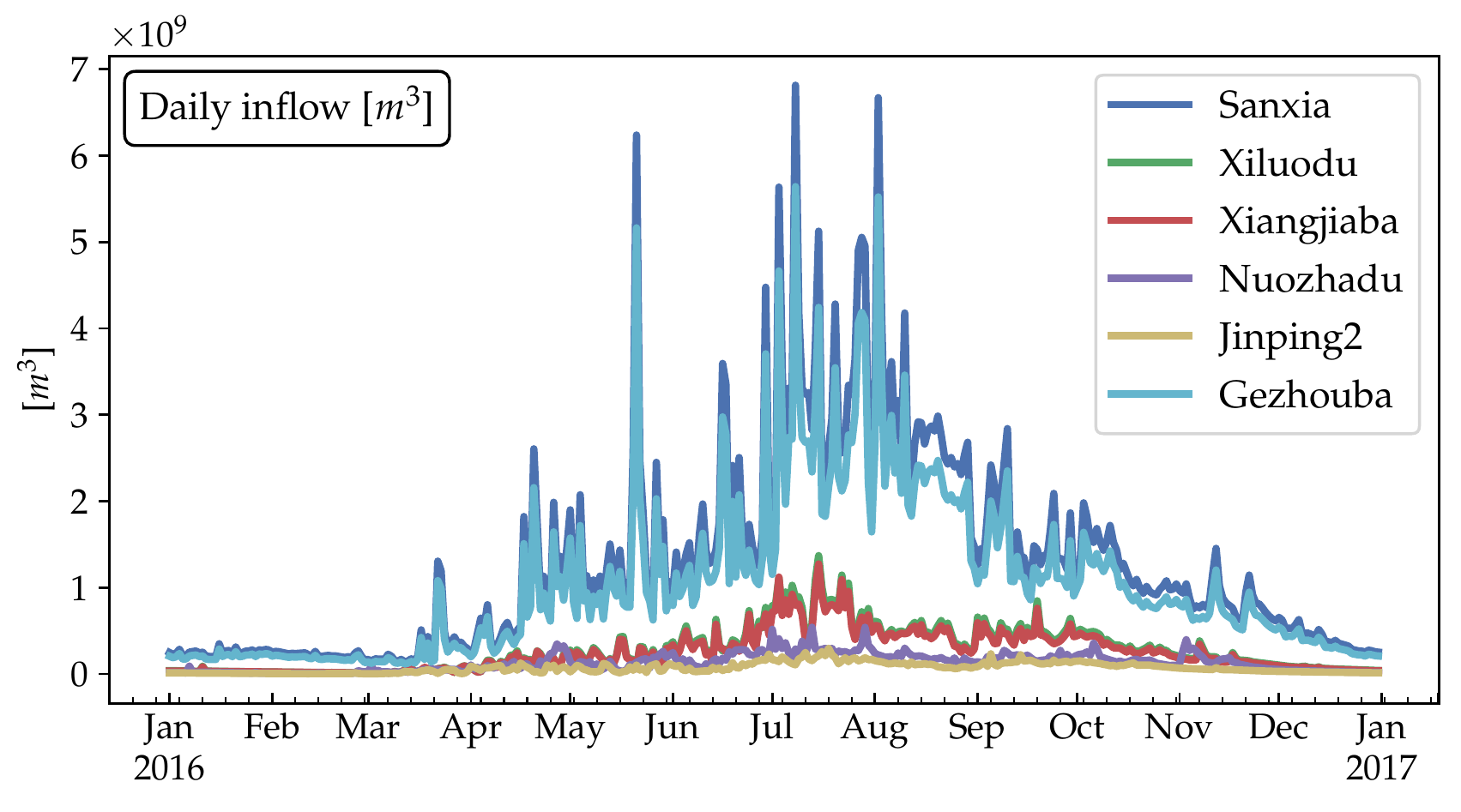}
	\includegraphics[width=\linewidth]{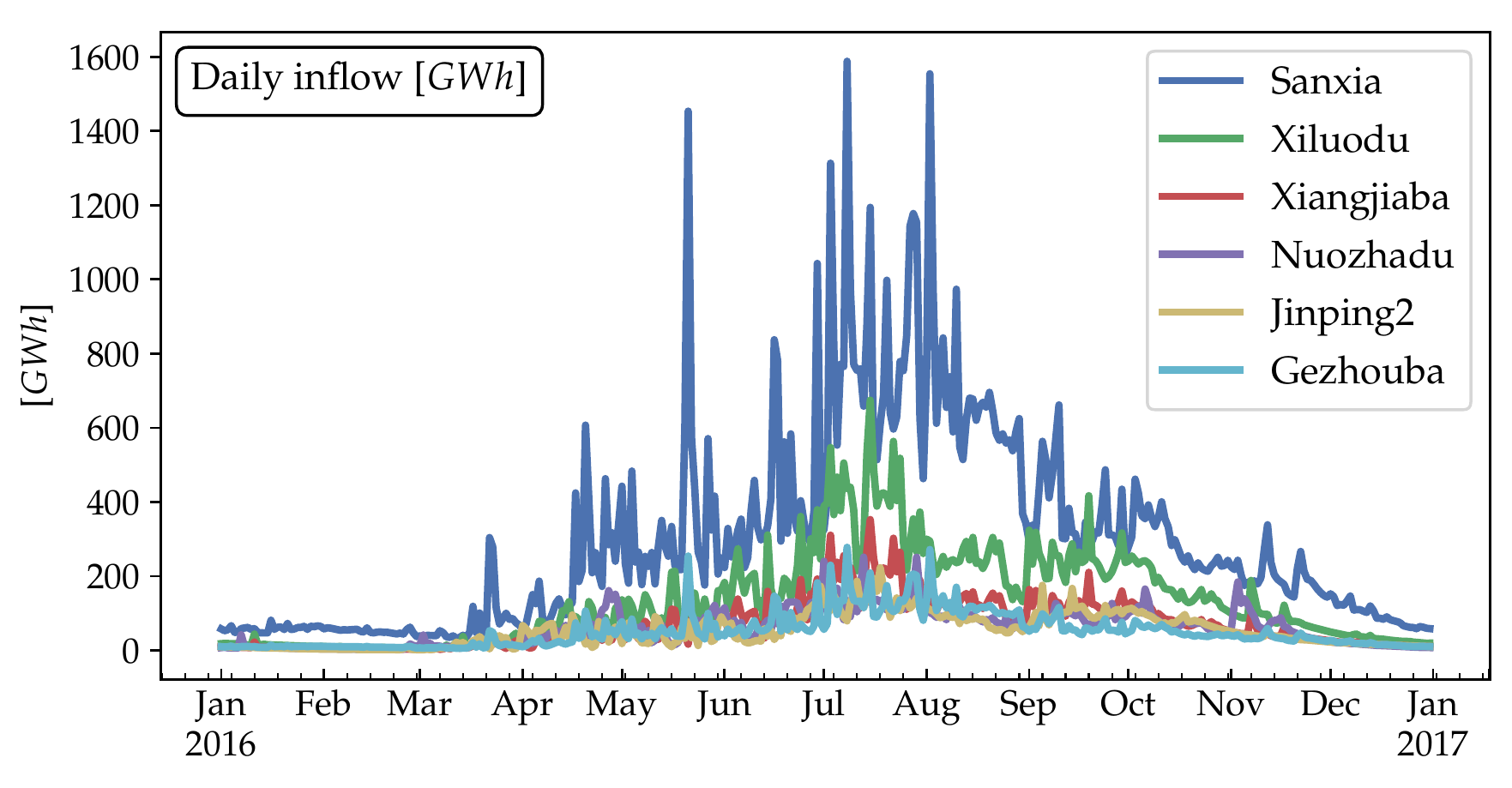}
	\caption{Daily reservoir inflow time series in 2016 in terms of water volume (top) and potential power production (bottom) for the 6 largest hydro stations. This dataset is freely available online at {http://doi.org/10.5281/zenodo.1471322} \cite{liu_hailiang_2018_1471322}.}
	\label{time_series}
\end{figure*}

In energy systems analyses, consistency of the feed-in data is vital. The surface runoff data in 1979-2016 from CFSR is deployed to produce the 38-year-long daily hydro inflow time series, as if the 41 hydro stations capacities remain the same. Shown in \fref{time_series}, two pairs of hydro stations have identical inflow time series, due to their proximity over the same river, but their potential power generation are different from each other, for they are distinct in head heights.

\section{Hydro station cascades}
\label{cascades}

One important character of hydro stations in China is that, they are usually part of a reservoir cascade, such as Three Gorges-Gezhouba, Xiluodu-Xiangjiaba,  Longyangxia-Laxiwa-Lijiaxia-Gongboxia-Qingtongxia \cite{SHANG201814,LU201556}. In such cascades, the reservoirs are chained along the same river, and the downstream reservoirs' inflow largely depends on their upstream stations' generator control or spillage. This is also accounted for in the model, assuming water flows into the downstream reservoir instantly. 

The cascades are represented as follows. Reservoirs are modeled as storage units with certain capacities (in terms of water volume) and daily inflows (charging) calculated as above. Only the first station of a cascade has a feed-in water flow. Its downstream stations' inflow follows its turbine control and spillage schedules. At the first dam, whenever water flows through the turbines, it generates power according to its head height and its downstream obtains the same amount of water as inflow. All reservoirs' initial capacity are set at 90\% of their total capacities, based on historical year-end values \cite{Almanac}, so reservoirs do not start empty. To summarize, water volume in a reservoir can be represented as:
\begin{equation}
R_t^{(n)} = R_{t-1}^{(n)} + G_t^{(n-1)}/ H^{(n-1)} + S_t^{(n-1)} - G_t^{(n)}/ H^{(n)} - S_t^{(n)}, \hspace{.5cm} n\in{2, 3,\dots,N}
\end{equation}
and
\begin{equation}
R_t^{(1)} = R_{t-1}^{(1)} + I_t^{(1)} - G_t^{(1)}/ H^{(1)} - S_t^{(1)},
\end{equation}
where $ R_t^{(n)} $ denotes the water volume of the $ n $th reservoir in the cascade at time $ t $, and $ G $ for power generation, $ H $ for head heights (power production per unit volume of water), $ S $ for spillage, $ I $ for water inflow.

\begin{figure*}[h!]
	\centering
	\includegraphics[width=\linewidth]{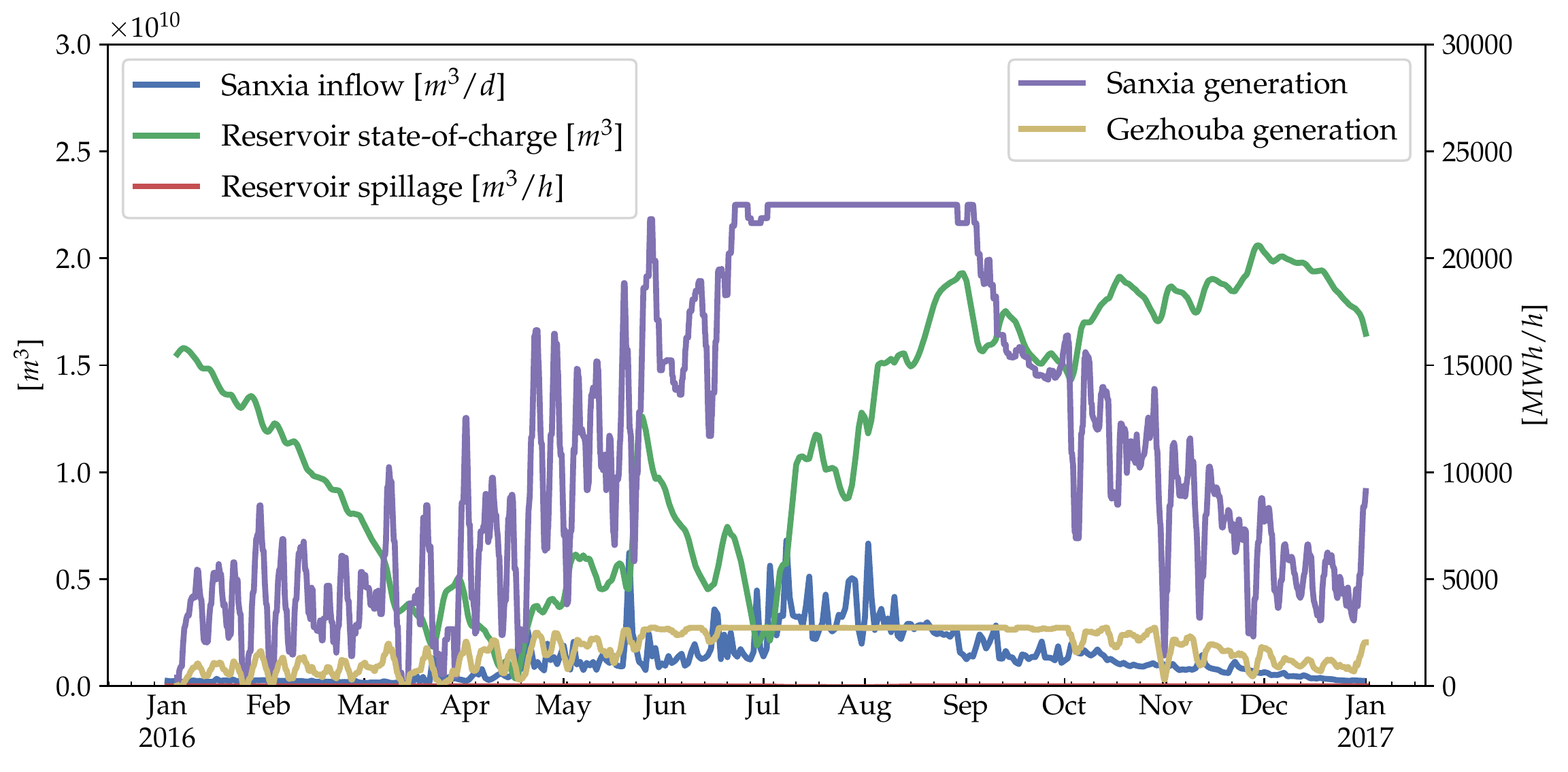}
	\caption{Visualization of reservoir inflow, state-of-charge, spillage, and power generation with weather year 2016 at \textit{Three Gorges (Sanxia)}, of which 38 km downstream lies another hydro station \textit{Gezhouba}. The former is equipped with a 39.3 billion $ m^3 $ reservoir, 181 m head height and generators totaling 22.5 GW, while the  latter 1.58 billion $ m^3 $, 47 m and 2.7 GW, respectively.  For better visualization, we only show 72-hour moving averages.}
	\label{hydro_ts}
\end{figure*}

\section{An application}

The inflow time series and the cascade model were integrated into a 31-node hourly-resolved techno-economic optimization of a wind and solar dominated Chinese power system, as the first application. Here, the operation of the reservoir hydro stations are optimized together with wind and solar capacity to minimize long term system investment cost, under ambitious carbon reduction targets. It is shown that the flexibility of hydro can complement the highly variable renewable generations and lower the cost of both transmission and renewable investments. Furthermore, its seasonality matches well with the summer peak load, unlike the renewables, thus can help reduce conventional power usage. A number of time series for the Three Gorges-Gezhouba cascade for weather year 2016 is shown in \fref{hydro_ts}. Detailed description and results are referred to the article \cite{Liu2018Oct}.

\FloatBarrier

\section{Acknowledgments}

The first author gratefully acknowledges the financial support from Idella Foundation Denmark and China Scholarship Council. G.B.A. and M.G. are partially funded by the RE-INVEST project (Renewable Energy Investment Strategies -- A two-dimensional interconnectivity approach), which is supported by Innovation Fund Denmark (6154-00022B). T.B. acknowledges funding from the Helmholtz Association, Germany under grant no. VH-NG-1352.  The responsibility for the contents lies solely with the authors.

\section{Bibliography}

\bibliographystyle{unsrtDOI}

\bibliography{hydro_paper}

\end{document}